\documentclass[graybox]{svmult}
\usepackage{mathrsfs}
\usepackage{amsmath}
\usepackage{mathptmx}       
\usepackage{helvet}         
\usepackage{courier}        
\usepackage{type1cm}        
\usepackage{makeidx}         
\usepackage{graphicx}        
\usepackage{multicol}        
\usepackage[bottom]{footmisc}
\usepackage{xcolor}          

\usepackage{bm}              

\usepackage{amssymb}

\DeclareFontFamily{OT1}{pzc}{}
\DeclareFontShape{OT1}{pzc}{m}{it}{<-> s * [1.200] pzcmi7t}{}
\DeclareMathAlphabet{\mathpzc}{OT1}{pzc}{m}{it}
\newcommand{\BEQ}{\begin{equation}}     
\newcommand{\BEA}{\begin{eqnarray}}
\newcommand{\BD}{\begin{displaymath}}
\newcommand{\EEQ}{\end{equation}}       
\newcommand{\EEA}{\end{eqnarray}}
\newcommand{\ED}{\end{displaymath}}
\newcommand{\vep}{\varepsilon}          
\renewcommand{\D}{{\rm d}}              
\newcommand{\II}{{\rm i}}               
\newcommand{\demi}{\frac{1}{2}}         
\newcommand{\wit}[1]{\widetilde{#1}}    
\newcommand{\wht}[1]{\widehat{#1}}      
\newcommand{\lap}[1]{\overline{#1}}     
\newcommand{\fR}{\mathfrak{R}}          
\newcommand{\bR}{\lap{\mathfrak{R}}}    

\renewcommand{\vec}[1]{\boldsymbol{#1}} 
\newcommand{\fns}{\footnotesize}        

\definecolor{gruen}{rgb}{0,0.625,0}     
\definecolor{rot}{rgb}{0.75,0,0}        
\definecolor{blau}{rgb}{0,0,0.75}       

\begin{document}

\title*{Schr\"odinger-invariance in phase-ordering kinetics}
\titlerunning{Phase-ordering kinetics}
\author{Stoimen Stoimenov and Malte Henkel}
\institute{Stoimen Stoimenov$^a$ \at $^a$Institute of Nuclear Research and Nuclear Energy, Bulgarian Academy of Sciences,
72 Tsarigradsko chaussee, Blvd., BG -- 1784 Sofia, Bulgaria
\and Malte Henkel$^{b,c}$ \at $^b$Laboratoire de Physique et Chimie Th\'eoriques {\small (CNRS UMR 7019)},
Universit\'e de Lorraine Nancy, B.P. 70239, F -- 54506 Vand{\oe}uvre-l\`es-Nancy Cedex, France\\
$^c$Centro de F\'{\i}sica T\'eorica e Computacional, Universidade de Lisboa,
P -- 1749-016 Lisboa, Portugal
}
\maketitle

\abstract{The generic shape of the single-time and two-time correlators in non-equilibrium phase-ordering kinetics with $\mathpzc{z}=2$  is obtained
from the co-variance of the four-point response functions. 
Their non-equilibrium scaling forms follow from a new non-equilibrium representation of the Schr\"odinger algebra.  
}

%

\section{Ageing in phase-ordering kinetics} \label{sec:age}

Phase-ordering kinetics \cite{Bray94a} has been studied since the 1960s. 
It concerns the growth of correlated microscopic clusters and as such is a paradigmatic example
of physical ageing \cite{Bray94a,Godr02,Cugl03,Puri09,Henk10,Taeu14,Cugl15}. 
In general, in a (classical) many-body system, ageing is brought about as follows \cite{Stru78}: 
prepare the system in an initially disordered, high-temperature state
and then quench it instantly to a low temperature $T$. Then fix the temperature and observe the dynamics. 
Phase-ordering kinetics is realised if that quench carries the system across
a phase-transition, which occurs at a critical temperature $T_c>0$, to some low temperature $T<T_c$. 
The microscopic inhomogeneity is described through a characteristic
time-dependent length-scale $\ell=\ell(t)$. We shall restrict attention to systems when this growth is algebraic, 
viz. $\ell(t)\sim t^{1/\mathpzc{z}}$ at large times, which
defines the critical exponent $\mathpzc{z}$. We shall be interested in a late-time and long-distance description 
when it is admissible to use a coarse-grained 
order-parameter $\phi(t,\vec{r})$, to be taken to be a continuous field. The system's behaviour is often analysed via 
the two-time {\em correlators} $C$ and two-time {\em responses} $R$, defined as 
\begin{subequations} \label{1.1}
\begin{align} \label{1.1a}
C(t,s;\vec{r}) &:= \left\langle \phi(t,\vec{r}) \phi(s,\vec{0}) \right\rangle =  F_C\left(\frac{t}{s},\frac{|\vec{r}|}{s^{1/\mathpzc{z}}}\right) \\
R(t,s;\vec{r}) &:= \left.\frac{\delta\bigl\langle \phi(t,\vec{r})\bigr\rangle}{\delta h(s,\vec{0})}\right|_{h=0}
                 =\left\langle \phi(t,\vec{r}) \wit{\phi}(s,\vec{0}) \right\rangle
                 = s^{-1-a} F_R\left(\frac{t}{s},\frac{|\vec{r}|}{s^{1/\mathpzc{z}}}\right) \label{1.1b}
\end{align}
\end{subequations}
where $h(s,\vec{r})$ is a symmetry-breaking external field conjugate to $\phi$. We shall always admit spatial translation- and
rotation-invariance such that $\vec{r}\mapsto r := |\vec{r}|$. Letting $t=s$ gives the {\em single-time correlator}: $C(s;r):=C(s,s;r)$ and
the {\em two-time autocorrelator} and {\em autoresponse} are defined as $C(t,s) := C(t,s;0)$ and $R(t,s) :=R(t,s;0)$. 
We shall review their determination from  dynamic symmetries. After this introduction to ageing, section~\ref{sec:nef} gives field-theoretic background
and our results are in section~\ref{sec:res}. 

\begin{figure}[tb]
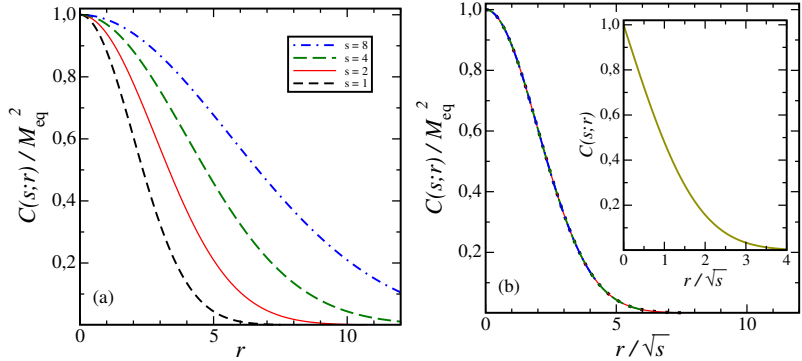

\begin{center}
\includegraphics[width=.45\hsize]{actes_LT16_phaseorder_spheriqueC1_vieil.eps}~
\includegraphics[width=.44\hsize]{actes_LT16_phaseorder_spheriqueC1_vieil-ska-inset.eps}~
\end{center}
\caption[fig1]{\small Ageing of the phase-ordering in the single-time correlator $C(s;r)$ in the (mean) spherical model in $d>2$  dimensions. 
The inset in panel (b) shows the form of the scaling function for a scalar order-parameter and the agreement with Porod's law.\label{fig1}}
\end{figure}

\begin{figure}[tb]
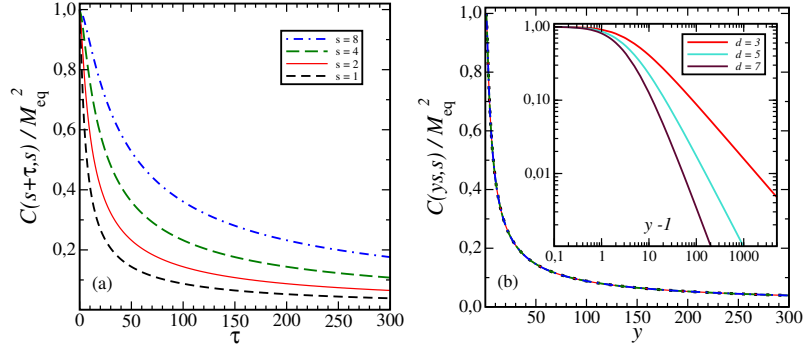

\begin{center}
\includegraphics[width=.45\hsize]{actes_LT16_phaseorder_spheriqueC2_vieil.eps}~
\includegraphics[width=.44\hsize]{actes_LT16_phaseorder_spheriqueC2_vieil-ska.eps}~
\end{center}
\caption[fig2]{\small Ageing of the phase-ordering in the two-time auto-correlator $C(t,s)$ in the $3D$ (mean) spherical model, see \cite{Henk25V}.
The inset in panel (b) shows the form of the scaling function for dimensions $d=[3,5,7]$ from top to bottom.  \label{fig2}}
\end{figure}

In figures~\ref{fig1} and~\ref{fig2} the further content of eqs.~(\ref{1.1}) for phase-ordering kinetics is illustrated. 
Figure~\ref{fig1}a shows the single-time auto-correlator $C(s;r)$, normalised by the equilibrium magnetisation $M_{\rm eq}^2$, for several times $s$. 
The first defining property of ageing \cite{Stru78,Henk10}, namely {\it slow dynamics}, 
appears since for increasing times $s$, the correlator decays more slowly. The second
property, {\it absence of time-translation-invariance}, is obvious since there is a distinct curve for each value of $s$. The third property, {\it dynamical scaling},
is displayed in figure~\ref{fig1}b, via the data collapse when the same data are replotted over against $r/\ell(s)$. We see that $\ell(s)\sim s^{1/2}$, in 
agreement with the expected value $\mathpzc{z}=2$ \cite{Bray94b} for phase-ordering when the dynamics of the order-parameter $\phi$ does not obey
any macroscopic conservation law (one speaks of {\em model-A-type dynamics}). The shape of the scaling function in figure~\ref{fig1}b reflects the fact that the
spherical model spins and their interfaces are quite `soft' such that $C(s;r)$ is rounded-off close to $r\approx 0$, as it occurs for vector-valued
order-parameters. For systems with `hard' interfaces, typical for scalar order-parameters, such as in the 
Glauber-Ising universality class, one rather observes a cusp at $r\approx 0$ as illustrated in the inset of figure~\ref{fig1}b. 
This cusp-like behaviour is known as {\em Porod's law} \cite{Poro51}. Experimentally, known examples for scalar model-A-dynamics occur in 
liquid crystals \cite{Alme21} and, with an anti-ferromagnetic order-parameter, in the binary alloy Cu$_3$Au \cite{Shan92}. 

Figure~\ref{fig2} presents the same kind of analysis for the two-time autocorrelator $C(ys,s)$.  The first two properties of ageing, slow dynamics and
absence of time-translation-invariance, are displayed in figure~\ref{fig2}a where $C$ is plotted over against the time different $\tau=t-s$, 
and the third property of the data collapse of dynamical scaling is shown in figure~\ref{fig2}b 
when the same data are replotted over against $y=t/s$. The inset further illustrates the
form of this $s$-independent scaling function $f_C(y)=C(ys,s)$. 
The overall form is (i) quite similar for all spatial dimensions $d$ and furthermore, (ii) for $y\gg 1$ one generically
finds a power-law $f_C(y)\sim y^{-\lambda_C/2}$, where $\lambda_C$ is the autocorrelation exponent \cite{Huse89}. 
The totality of the observations from figures~\ref{fig1},\ref{fig2} 
can be condensed into the single scaling form
quoted in (\ref{1.1a}). The scaling function $F_C$ is expected to be universal \cite{Bray90,Bray94a}, 
which means that its form should be independent of `microscopic details' such as the lattice structure, the
precise form of the interactions or the value of $T$. It does depend, 
however, on the spatial dimension $d$ and on the nature of the order-parameter (e.g. its symmetries). 

Similar observations can also be made for the response function and lead to (\ref{1.1b}), where $a$ is an ageing exponent. For the auto-reponse scaling
function, there is a power-law for $y\gg 1$, viz. $f_R(y)=s^{1+a}R(ys,s)\sim y^{-\lambda_R/2}$, 
where $\lambda_R$ is the $d$-dependent autoresponse exponent. The scaling function
$F_R$ is expected universal as well and for spatially short-ranged initial correlations, 
one finds $\lambda = \lambda_C=\lambda_R$. In (\ref{1.1b}) we also anticipate
Janssen-de Dominicis non-equilibrium field theory which allows to 
rewrite the response functions formally as a correlator with a so-called {\em response scaling operator} \cite{Jans76,Domi76}. 

Rather than studying any specific theory of phase-ordering kinetics, 
we inquire about generic deteriminations of the universal scaling functions $F_{C,R}$. Symmetry arguments
are an obvious candidate. They should lead to an understanding of the scaling forms (\ref{1.1}) and the properties of the scaling functions $F_{C,R}$. 
Since the dynamical exponent $\mathpzc{z}=2$, a promising candidate for a larger set of dynamical symmetries might appear to be 
{\em Schr\"odinger-transformations}, already discovered by Jacobi and by Lie in the 19$^{\rm th}$ century, and defined as
\BEQ \label{1.2}
t \mapsto t' = \frac{\alpha t +\beta}{\gamma t + \delta} \;\; , \;\;
\vec{r} \mapsto \vec{r}' = \frac{\mathscr{R}\vec{r} + \vec{v}t + \vec{a}}{\gamma t +\delta} \;\; ; \;\;
\alpha\delta - \beta\gamma =1
\EEQ
where $\mathscr{R}\in\mbox{\sl SO}(d)$ is a rotation matrix and $\vec{v},\vec{a}\in\mathbb{R}^d$ are vectors. For a historical review, see \cite{Duva24}. 
Certainly, time-translations are included therein and further considerations will be needed to render the usual Schr\"odinger-transformations (\ref{1.2}) 
applicable to non-equilibrium ageing. This will be described in the next section.

\section{Background: field-theory \& dynamical symmetry} \label{sec:nef}

The forthcoming discussion of the scaling form (\ref{1.1}), for phase-ordering kinetics, 
in section~\ref{sec:res} will rely on field-theoretic methods 
and a new adaptation of Schr\"odinger-invariance \cite{Henk25e}. 
We refer to the detailed exposition of these techniques in \cite{Henk25V} 
(done there mainly for non-equilibrium critical dynamics after a quench onto $T=T_c$) 
and shall limit ourselves to indicating the necessary differences.

{\bf 1.} Physical ageing must be set into the context of non-equilibrium continuum field-theory \cite{Domi76,Jans76,Taeu14}.
In principle, one calculates the average of an observable $\mathscr{A}$ via a functional integral
$\bigl\langle \mathscr{A}\bigr\rangle = \int \mathscr{D}\phi\mathscr{D}\wit{\phi}\; \mathscr{A}[\phi]\, e^{-{\cal J}[\phi,\wit{\phi}]}$. 
For phase-ordering kinetics, with non-conserved model-A-type dynamics of the order-parameter, 
one has ${\cal J}[\phi,\wit{\phi}\,] = {\cal J}_0[\phi,\wit{\phi}\,] + {\cal J}_b[\wit{\phi}\,] $ for the Janssen-de Dominicis action where 
\begin{subequations} \label{actionJdD}
\begin{align}
{\cal J}_0[\phi,\wit{\phi}\,] &= \int \!\D t\D\vec{r}\: \left( \wit{\phi} \left( \partial_t - \Delta_{\vec{r}} - V'[\phi]\right)\phi  \right) \\
{\cal J}_b[\wit{\phi}\,] &= -\demi \int_{\mathbb{R}^{2d}} \!\D\vec{R}\D\vec{R}'\: \wit{\phi}(0,\vec{R}) C_0\bigl(\vec{R}-\vec{R}'\bigr) \wit{\phi}(0,\vec{R}') 
\label{actionJdDc} 
\end{align}
\end{subequations}
with the interaction $V'[\phi]$ and the spatial laplacian $\Delta_{\vec{r}}$ (and the usual re-scalings). 
Any noise comes only from the spatially short-ranged `initial'
correlator $C_0\bigl(\vec{R}\bigr)$. The {\em deterministic action} ${\cal J}_0[\phi,\wit{\phi}\,]$ gives rise to the 
{\em deterministic average} 
$\bigl\langle \mathscr{A}\bigr\rangle_0 = \int \mathscr{D}\phi\mathscr{D}\wit{\phi}\; \mathscr{A}[\phi]\, e^{-{\cal J}_0[\phi,\wit{\phi}]}$. 
This is important because either causality \cite{Jans76,Cala05,Taeu14} 
or the combination of Galilei- and spatial translation-invariance \cite{Pico04} of ${\cal J}_0[\phi,\wit{\phi}\,]$
imply the Barg\-man su\-per\-se\-lec\-tion rules $\left\langle \overbrace{~\phi \cdots \phi~}^{\mbox{\rm ~~$n$ times~~}}
             \overbrace{ ~\wit{\phi} \cdots \wit{\phi}~}^{\mbox{\rm ~~$m$ times~~}}\right\rangle_0 \sim \delta_{n,m}$ \cite{Barg54}. 
Non-vanishing deterministic averages must have an equal number of order-parameters $\phi$
and conjugate response operators $\wit{\phi}$. Examples are two-point response functions (see \ref{1.1b}))
$R=\bigl\langle\phi\wit{\phi}\bigr\rangle=\bigl\langle\phi\wit{\phi}\bigr\rangle_0$
or four-point responses $\bigl\langle\phi\phi\wit{\phi}\wit{\phi}\bigr\rangle=\bigl\langle\phi\phi\wit{\phi}\wit{\phi}\bigr\rangle_0$. 
On the other hand, a correlator $\left\langle\phi\phi\right\rangle$ must be obtained from a four-point response function 
\BEA
\lefteqn{\hspace{-0.9cm}C(t,s;r)
= \bigl\langle \phi(t,\vec{r}+\vec{r}_0)\phi(s,\vec{r}_0) \bigr\rangle 
= \int \mathscr{D}\phi\mathscr{D}\wit{\phi}\; \phi(t,\vec{r}+\vec{r}_0)\phi(s,\vec{r}_0)\, e^{-{\cal J}_0[\phi,\wit{\phi}]-{\cal J}_b[\wit{\phi}]}~~} \nonumber \\
&=& \bigl\langle \phi(t,\vec{r}+\vec{r}_0)\phi(s,\vec{r}_0) e^{-{\cal J}_b[\wit{\phi}]} \bigr\rangle_0 \nonumber \\
&=& \demi \int_{\mathbb{R}^{2d}} \!\!\!\D\vec{R}\D\vec{R}'\: C_0\bigl(\vec{R}-\vec{R}'\bigr)
\left\langle \phi(t,\vec{r}+\vec{r}_0) \phi(s,\vec{r}_0) \wit{\phi}(\vep,\vec{R}) \wit{\phi}(\vep',\vec{R}') \right\rangle_0
\label{corrFT}
\EEA
by expansion to all orders of the exponential $e^{-{\cal J}_b[\wit{\phi}]}$ of which a single contribution will remain \cite{Pico04,Henk10}. 
This replaces \cite[eq. (4)]{Henk25V} in the case of phase-ordering and will serve as our starting point. 
In (\ref{corrFT}) $\vep,\vep'$ are `initial' time-scales, to be fixed later.

{\bf 2.} The Schr\"odinger group is known to be the maximal finite-dimensional symmetry of the free Schr\"odinger equation 
$\mathscr{S}\phi=\bigl(2{\cal M}\partial_t - \Delta_{\vec{r}}\bigr)\phi=0$ 
in the sense that it maps any solution of that equation onto another solution. This implies
that the deterministic action ${\cal J}_0[\phi,\wit{\phi}]$ is Schr\"odinger-invariant, 
as shown explicitly for free fields \cite{Henk03a}
or the $(1+1)D$ Calogero model \cite{Shim21}. Since the Lie algebra which follows from (\ref{1.2}) 
is not semi-simple, its representations must be projective and
we refer to \cite{Henk25V} for the explicit generators. For the standard representation of the Schr\"odinger Lie algebra, 
the order-parameter $\phi$ is characterised by a scaling dimension $\delta$. 
Then the hypothesis of Schr\"odinger-covariance leads to the two-point function 
($\mathscr{R}_0$ is a normalisation constant) \cite{Henk94}
\BEA
\lefteqn{R(t_a,t_b;r) = \left\langle \phi_a(t_a,\vec{r}) \wit{\phi}_b(t_b,\vec{0})\right\rangle_0} \nonumber \\
&=&  \mathscr{R}_0\, \delta({\cal M}_a + \wit{\cal M}_b)\,\delta_{\delta_a,\wit{\delta}_b}\: \Theta(t_a-t_b) 
\bigl( t_a - t_b\bigr)^{-2\delta_a} \exp\left[ - \frac{{\cal M}_a}{2} \frac{\vec{r}^2}{t_a-t_b} \right]
\label{2points}
\EEA
Response operators $\wit{\phi}$ have negative masses $\wit{\cal M}_b = \wit{\cal M}=-{\cal M}=-{\cal M}_a<0$. 
In addition, there is the constraint $\wit{\delta}=\delta$ 
between the scaling dimension of the response operator and the order-parameter. 
The generic Schr\"odinger-covartiant four-point function is \cite{Golk14,Shim21,Volo09}
\BEA
\lefteqn{\left\langle \phi(t,\vec{r})\phi(s,\vec{0}) \wit{\phi}(0,\vec{R})\wit{\phi}(0,\vec{R}')\right\rangle_0} \nonumber \\
&\simeq& \bigl(ts\bigr)^{-2\delta} \exp\left[-\frac{\cal M}{2}\frac{\vec{r}^2}{t-s}-{\cal M}\frac{\fR^2+\bR^2}{s}\right] 
\mathscr{F}^{(2)}\left(\frac{\fR}{s^{1/2}}+\frac{1}{y}\frac{\vec{r}}{s^{1/2}},\frac{\bR}{s^{1/2}}\right)  
\label{eq:2.8}
\EEA
with the new space variables $\fR = \demi\bigl( \vec{R} + \vec{R}'\bigr)$ and $\bR = \demi\bigl( \vec{R} - \vec{R}'\bigr)$ and $y=t/s$. 
We shall also need the so-called `pairwise equal-time case' when $t=s>0$ and \cite{Shim21} 
\BEA
\lefteqn{\left\langle \phi(s,\vec{r})\phi(s,\vec{0}) \wit{\phi}(0,\vec{R})\wit{\phi}(0,\vec{R}')\right\rangle_0} \nonumber \\
&\simeq& s^{-4\delta}\exp\left[-{\cal M}\frac{\fR^2+\bR^2}{s}+{\cal M}\frac{\vec{r}\cdot\fR}{s}\right]
\mathscr{F}^{(1)}\left(\frac{\vec{r}\cdot\bR}{s}\right)
\label{eq:2.9}
\EEA 
where $\mathscr{F}^{(1,2)}$ are undetermined (differentiable) functions of one/two arguments, 
respectively, which are not fixed by Schr\"odinger-covariance alone. 
These expressions were directly written in the scaling limit 
\begin{subequations} \label{gl:skal}
\BEQ
s\to\infty \;\; , \;\; \tau = t-s = (y-1)s \to \infty \;\; , \;\; \vec{r}, \vec{R}, \vec{R}' \to \infty
\EEQ
such that the following quantities are kept finite
\BEQ
y = \frac{t}{s} > 1\;\; , \;\; \frac{\vec{r}}{s^{1/2}} \;\; , \;\; \frac{\vec{R}}{s^{1/2}} \;\; , \;\; \frac{\vec{R}'}{s^{1/2}}
\EEQ
\end{subequations}
In both (\ref{eq:2.8},\ref{eq:2.9}), the time-scale `$0$' of the response operators $\wit{\phi}$ is meant as a short-hand
for an `initial' time-scale $\vep \ll s,t$, to be specified below. 
Finally, for applications which involve finite-size effects, one may reuse (\ref{eq:2.8}) but with
the finite-size scaling function 
$\mathscr{F}^{(2,N)}\left(\frac{\fR}{s^{1/2}}+\frac{1}{y}\frac{\vec{r}}{s^{1/2}},\frac{\bR}{s^{1/2}},\frac{t^{1/2}}{N}\right)$ 
for $t\gg s$ \cite{Henk25e}.

{\bf 3.} The expressions (\ref{eq:2.8},\ref{eq:2.9}) for the two- and four-point response are brought out-of-equilibrium  by the following  

\noindent
{\bf Postulate:} \cite{Henk25,Henk25c} {\it The Lie algebra generator $X_n^{\rm equi}$ 
of a time-space symmetry of an equilibrium system becomes a symmetry
out-of-equilibrium by the change of representation}
\BEQ \label{gl:hyp}
X_n^{\rm equi} \mapsto X_n = e^{\xi \ln t} X_n^{\rm equi} e^{-\xi \ln t}
\EEQ
{\it where $\xi$ is a dimensionless parameter whose value contributes to characterise the scaling operator $\phi$ on which $X_n$ acts.}

For critical dynamics, at $T=T_c$, this is suggestive since one may consider 
this as a generalisation of known equilibrium dynamical symmetries \cite{Card85}. 
In that case, there a numerous practical examples, reviewed in \cite{Henk25V}, 
which suggest that the method might work successfully. 
For phase-ordering at $T<T_c$, however, despite the well-established non-equilibrium dynamical scaling 
\cite{Bray90,Bray94a}, it is less obvious that our postulate should work
and does require separate testing \cite{Henk25e}.  

Formally, when applied to the dilatation generator $X_0^{\rm equi}=-t\partial_t-\delta$ this leads to 
\begin{subequations} \label{gl:Xgen}
\BEQ \label{gl:X0gen}
X_0^{\rm equi} \mapsto X_0 = -t\partial_t - \frac{1}{\mathpzc{z}}r\partial_r - \bigl(\delta - \xi\bigr)
\EEQ
which means that one has an effective scaling dimension $\delta_{\rm eff} = \delta-\xi$. 
The time-translation generator $X_{-1}^{\rm equi}=-\partial_t$ turns into 
\BEQ \label{gl:X-1gen}
X_{-1}^{\rm equi} \mapsto X_{-1} = -\partial_t + \frac{\xi}{t}
\EEQ
\end{subequations}
which makes the result of an application of $X_{-1}$ appear non-trivial. 
Significantly, in this new representation the scaling operators become 
$\Phi(t) = t^{\xi} \phi(t) = e^{\xi \ln t}\phi(t)$ which will be identified as the `physical' ones. 
The above equilibrium response functions, found from covariance 
under the standard representation of the Schr\"odinger Lie algebra \cite{Henk25V}, now read 
(spatial arguments are suppressed for clarity)
\BEA
\bigl\langle \phi_a(t_a)\wit{\phi}_b(t_b)\bigr\rangle_0
&\mapsto & t_a^{\xi_a} t_b^{\wit{\xi}_b} \: \bigl\langle \phi_a(t_a)\wit{\phi}_b(t_b)\bigr\rangle_0
\nonumber \\
\bigl\langle \phi_a(t_a)\phi_b(t_b)\wit{\phi}_c(t_c)\wit{\phi}_d(t_d)\bigr\rangle_0
&\mapsto &
t_a^{\xi_a} t_b^{\xi_b} t_c^{\wit{\xi}_c} t_d^{\wit{\xi}_d} \: \bigl\langle \phi_a(t_a) \phi_b(t_b)\wit{\phi}_c(t_c)\wit{\phi}_d(t_d)\bigr\rangle_0
\label{reponseHE}
\EEA
Now, we characterise a {\em non-equilibrium} scaling operator $\phi$ 
by a pair of scaling dimensions $(\delta,\xi)$ and a non-equilibrium response operator
$\wit{\phi}$ by a pair $(\wit{\delta},\wit{\xi})$. 
The Bargman rule with $n=m=1$ implies $\delta=\wit{\delta}$ but $\xi$ and $\wit{\xi}$ remain independent.

Finally, for the two-time autocorrelator 
$C(t,s)=\langle \phi_1\phi_2\rangle=\langle \phi(t)\phi(s)\rangle$ the scaling operator
identity $\phi_1=\phi_2=\phi$ implies for the scaling dimensions $\delta_1=\delta_2=\delta$ and $\xi_1=\xi_2=\xi$. 
This produces the exponent relations \cite{Henk25c} 
\BEQ \label{gl:Cphaseord}
\frac{\lambda}{2} = 2\delta - \xi \;\; , \;\; \delta = \xi
\EEQ

{\bf 4.} Once correlated domains have formed, 
the effective equation of motion is no longer the one derived from the action (\ref{actionJdD}) which becomes unstable rapidly \cite{Bray94a}
but will rather take an effective form 
$\mathscr{S}^{\rm equi}\phi=\left( \partial_t - \frac{1}{2{\cal M}} \Delta_{\vec{r}} \right) \phi(t,\vec{r}) = g \phi^3(t,\vec{r})$. 
The plausibility of this form is argued as follows \cite{Henk25c}: 
\begin{enumerate}
\item a term linear in $\phi(t,\vec{r})$ on its right-hand-side would break dynamical scaling
\item a term quadratic in $\phi(t,\vec{r})$ breaks the global spin-reversal-invariance {$\phi\mapsto -\phi$}
\item a term cubic in $\phi(t,\vec{r})$ is the lowest-order term which may appear 
\item thermal noise will merely lead to corrections to scaling  
\item the exponent $\mathpzc{z}=2$ \cite{Bray94b} for short-ranged model-A-type dynamics
\end{enumerate}

Our postulate implies the modified form of the Schr\"odinger operator $\mathscr{S}^{\rm equi}=\partial_t - \frac{1}{2{\cal M}} \Delta_{\vec{r}} $ 
\BEQ \label{gl:phi-eff2}
{\mathscr{S}} = e^{\xi \ln t} \mathscr{S}^{\rm equi} e^{-\xi \ln t} = \partial_t - \frac{\xi}{t} - \frac{1}{2{\cal M}} \Delta_{\vec{r}}
\EEQ
and contains an additional $1/t$-potential which is well-known from the literature \cite{Oono88,Maze06}.
Because of $\left( t^{\xi}\, \mathscr{S}^{\rm equi}\, t^{-\xi} \right) \left( t^{\xi}\, \phi \right) = g\, t^{\xi} \left( t^{-\xi}\, {\Phi}\, \right)^{3}$
we find
\BEQ \label{gl:phi-eff3}
{\mathscr{S}}\,{\Phi} = \left( \partial_t - \frac{\xi}{t} - \frac{1}{2{\cal M}} \Delta_{\vec{r}} \right) {\Phi} = g\, t^{-2\xi}\, {\Phi}^{\,3}
\EEQ
For phase-ordering kinetics, (\ref{gl:Cphaseord}) implies that $\delta_{\rm eff}=\delta-\xi=0$ such that ${\Phi}$ is dimensionless, such that
the long-time behaviour of (\ref{gl:phi-eff3}) is governed by the explicit $t$-dependence.
The $1/t$-potential will for large times dominate over against the non-linear term, when the {\bf criterion} \cite{Henk25c,Henk25e}
\BEQ \label{gl:lambda-crit}
2\xi > 1 ~~\Longleftrightarrow~~ \lambda > 1
\EEQ
is satisfied. For its validity in models, recall the well-known auto-correlation bound 
$\lambda\geq d/2$ \cite{Fish88a,Yeun96a}. Hence for $d>2$, the criterion (\ref{gl:lambda-crit}) is satisfied. For $d=2$, one has
typically $\lambda\approx 1.25 >1$ (see \cite{Henk10} and refs. therein) and (\ref{gl:lambda-crit}) is satisfied as well.
Although the effective equation of motion of phase-ordering need not be Schr\"odinger-invariant,
we may use the Schr\"odinger symmetry of the linear part of (\ref{gl:phi-eff3}), with the additional $1/t$-potential, to deduce its long-time behaviour. 
Of course, this linearised equation cannot be used for a first-principles
calculation of exponents such as $\lambda, a, \ldots$ for which the full equation of motion must be used \cite{Maze06}. 

\section{Results} \label{sec:res}

We shall concentrate on the analysis of the correlators by using (\ref{corrFT}) as the starting point. Concerning the two-time response function,
we merely mention the well-known fact that Schr\"odinger-covariance does reproduce 
$f_R(y)\sim y^{-\lambda_R/2}$ for $y\gg 1$ and that $\lambda:=\lambda_R=\lambda_C$ \cite{Henk06,Henk25c}. 
In addition, in phase-ordering kinetics, in all known models one has $\delta=\xi=-\wit{\xi}$.

{\bf 1.} We begin with the two-time auto-correlator $C(ys,s)$. 
Combining (\ref{corrFT},\ref{eq:2.8},\ref{reponseHE}) and setting $\vec{r}=\vec{0}$, we find
\BEA
\lefteqn{C(ys,s;\vec{0}) = \!\int_{\mathbb{R}^{2d}} \!\!\D\fR\D\bR\: C_0\bigl(2\bR\bigr)\, 
y^{-2\delta+\xi} s^{2\xi-4\delta} \vep^{2\wit{\xi}}\:e^{-\frac{{\cal M}}{s}\bigl(\fR^2+\bR^2\bigr)}} \nonumber \\
& & \hspace{1.0cm} \times ~~
\mathscr{F}^{(2)}\left(\fR\frac{(s-1)^{1/2}}{s},\bR\frac{(s-1)^{1/2}}{s}\right)
\label{gl:3.2} \\
&\simeq& y^{-2\delta+\xi}\, s^{2(\xi-\delta)}\,s^{-2\delta+d} \vep^{2\wit{\xi}}
\underbrace{\int_{\mathbb{R}^{2d}} \!\D \vec{U}\D\lap{\vec{U}}\: C_0\bigl(2 \lap{\vec{U}} s^{1/2}\bigr) 
e^{-{\cal M}\bigl(\vec{U}^2+\lap{\vec{U}}^2\bigr)}\, \mathscr{F}^{(2)}\bigl(\vec{U}, \lap{\vec{U}}\bigr)}_{\to~ \mathscr{C}^{(2)}_{\infty}} 
\nonumber 
\EEA
where in the second line, we first let $s\gg 1$ and then changed the integration variables.
In what follows, we shall always assume that the initial correlator $C_0\bigl(\vec{R}\bigr)$ as well as
the scaling function $\mathscr{F}^{(2)}\bigl(\vec{U},\lap{\vec{U}}\bigr)$ 
are such that in the indicated limit of large waiting times $s\gg 1$ the integral tends towards a finite, non-vanishing constant $\mathscr{C}^{(2)}_{\infty}$. 
Furthermore, as inspired by the studies in \cite{Zipp00,Andr06}, we admit that the `initial' time-scale at the beginning
of the scaling regime is related to the waiting time $s$ as
\BEQ \label{gl:3.4}
\vep \simeq \vep_0\,  s^{\zeta_p}
\EEQ
where $\zeta_p$ is a new exponent supposed to describe the beginning of the scaling regime. With these assumptions,
the leading large-time behaviour (\ref{gl:3.2}) of the two-time auto-correlator becomes
\BEQ \label{gl:3.5}
C(ys,s) = y^{-(2\delta-\xi)}\, s^{2(\xi-\delta)}\,s^{d-2\delta+2\zeta_p\wit{\xi}}\: \vep_0\, \mathscr{C}^{(2)}_{\infty}
\EEQ
This already reproduces (i) the algebraic behaviour (\ref{1.1a}) of the two-time correlator for $y=t/s\gg 1$, and 
(ii) also shows that $\frac{\lambda_C}{2}=2\delta-\xi=\frac{\lambda_R}{2}$, as expected.  
Since for phase-ordering kinetics, one has (\ref{gl:Cphaseord}) and $\delta=\xi=-\wit{\xi}$.
The scaling (\ref{gl:3.5}) becomes $s$-independent, as expected from (\ref{1.1a}), if we have the new scaling relation \cite{Henk25e} 
\BEQ \label{gl:3.6}
2\delta = \frac{d}{1+\zeta_p} = \lambda
\EEQ
This scaling relation is distinct with respect to non-equilibrium critical dynamics. 
It underscores the non-trivial nature of the auto-correlation exponent $\lambda$.

For the passage exponent, one has obviously $\zeta_p\geq 0$ and also $\zeta_p\leq 1$ 
since the ageing regime cannot start later than at the waiting time $s$ itself. 
This reproduces the well-known bounds $\frac{d}{2} \leq \lambda \leq d$, 
from the literature \cite{Fish88a,Yeun96a}.

{\bf 2.} Now, we set $t=s$, combine  (\ref{corrFT},\ref{eq:2.9},\ref{reponseHE}) and have the single-time correlator 
\BEA
\lefteqn{ C(s;r) = \int_{\mathbb{R}^{2d}}\!\D\fR\D\bR\: C_0\bigl(2\bR\bigr)\: 
e^{-\frac{{\cal M}}{s}\bigl(\fR^2+\bR^2\bigr)+\frac{\cal M}{s}r\,\fR}\, s^{2\xi-4\delta}\vep^{2\wit{\xi}}
\mathscr{F}^{(1)}\left( \frac{\vec{r}\cdot\bR}{s} \right) }
\nonumber \\
&=& e^{-\frac{\cal M}{4}\frac{r^2}{s}}\, s^{2\xi-4\delta+\frac{d}{2}-2\xi\zeta_p}
\int_{\mathbb{R}^d}\!\D\bR\: C_0\bigl(2\bR\bigr)\: 
e^{-\frac{\cal M}{s} \bR^2}\, \mathscr{F}^{(1)}\left( \frac{\vec{r}}{s^{1/2}}\cdot\frac{\bR}{s^{1/2}} \right)
\nonumber \\
&=& e^{-\frac{\cal M}{4}\frac{r^2}{s}}\:
\int_{\mathbb{R}^d}\!\D\lap{\vec{U}}\: C_0\bigl(2\lap{\vec{U}}\, s^{1/2}\bigr)\: 
e^{-{\cal M}\lap{\vec{U}}^2}\, \mathscr{F}^{(1)}\left( \frac{\vec{r}}{s^{1/2}}\cdot \lap{\vec{U}} \right)
\label{gl:C1corr}
\EEA
Herein, we introduced the `initial' time estimate (\ref{gl:3.4}), 
completed a square in the $\fR$-integration, and applied again he scaling relation (\ref{gl:3.6}). 
If the same kind of limit as before exists and is finite, we have
again reproduce the scaling form (\ref{1.1}), now for $t=s$ and identify the scaling function $F_C$ with the natural scaling variable $r/\sqrt{s\,}$.  

An explicit computation of the function $F_C(1,r/\sqrt{s\,}\,)$ 
must await stronger information on $\mathscr{F}^{(1)}$ than is currently available.
If a limited analytic expansion of $\mathscr{F}^{(1)}$ for small arguments is possible, we would find an expansion of $C(s;r)$ for small $r$
\BEQ \label{gl:C1-21}
C\bigl(s;|\vec{r}|\bigr) \simeq \exp\left[ {-\frac{\cal M}{4}\frac{r^2}{s}}\right]
\int_{\mathbb{R}^d}\!\D\lap{\vec{U}}\: C_0\bigl(2\lap{\vec{U}} s^{1/2}\bigr)\: e^{-{\cal M}\lap{\vec{U}}^2}\,
\left( \mathscr{F}_0^{(1)} + \mathscr{F}_1^{(1)}\, \lap{\vec{U}}\cdot \frac{\vec{r}}{s^{1/2}} + \ldots \right)
\EEQ
If the corresponding integrals have finite limits for $s\gg 1$ and if $\mathscr{F}_1^{(1)}<0$, 
this would reproduce the typical small-distance behaviour 
$C(s;|\vec{r}|)\simeq C_0 - C_1 \frac{|\vec{r}|}{s^{1/2}} +\ldots$ for a scalar order-parameter, with a cusp at $r=0$. 
This is illustrated in the inset of figure~\ref{fig1}b and the observed linear behaviour is predicted by Porod's law \cite{Poro51,Bray94a}. A recent simulation illustrates this in \cite{Chris19}, and for a classic example see \cite[fig. 14]{Bray94a}.

Remarkably, single-time and two-time correlators are treated on the same conceptual basis,
namely the covariance of the four-point response function $\langle\phi\phi\wit{\phi}\wit{\phi}\rangle$.

{\bf 3.} We use (\ref{gl:C1corr}) in the definition of the {\em structure factor} and find
\BEA
\wht{S}(s;\vec{q}) &:=& \int_{\mathbb{R}^d} \!\D\vec{r}\: e^{-\II \vec{q}\cdot\vec{r}}\: C(s;\vec{r}) 
= s^{d/2} \mathpzc{g}\left( \vec{q} s^{1/2} \right) \:=\: \ell(s)^d \bar{\mathpzc{g}}\bigl( \vec{q} \ell(s) \bigr)
\EEA
the required scaling form \cite{Bray94a}, 
with scaling functions $\mathpzc{g}$ or $\bar{\mathpzc{g}}$, 
and the length scale $\ell=\ell(s)\sim s^{1/2}$. This is based on the same assumptions on $C_0\bigl(\vec{R}\bigr)$ and $\mathscr{F}^{(1)}$ as before.

In the limit  $|\vec{q}|\to\infty$, this should be compatible with Porod's law \cite{Poro51,Bray94a}. 
It is one of the central ingredients
in the derivation of $\mathpzc{z}=2$ for model-A-type dynamics in phase ordering \cite{Bray94b}. 
Indeed, on the basis of the expansion carried out in (\ref{gl:C1-21}), 
it can be shown that for large momenta $\vec{Q}=\vec{q} s^{1/2}\to \infty$, 
one obtains $\mathpzc{g}(\vec{Q}) \sim |\vec{Q}|^{-d-1}$ which is indeed the form in which Porod's law is usually stated.

\begin{figure}[tb]
\sidecaption
\includegraphics[width=0.3\hsize]{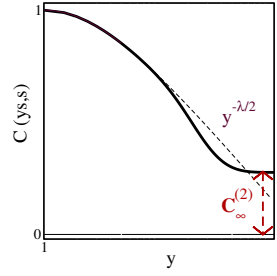}
\caption[fig3]{\small Autocorrelator $C(ys,s)$ in a fully finite system.  \label{fig3}}
\end{figure}

{\bf 4.} We now consider a fully finite system, say in a hypercubic geometry with a side of linear length $N$. 
A typical  autocorrelator $C(ys,s)$ is shown in figure~\ref{fig3}. 
For large system sizes $N\to\infty$, one recovers the behaviour of the infinite-size
system, with its power-law decay $f_C(y)\sim y^{-\lambda/2}$ (dashed line). 
For $N$ small, the correlator decreases with $y$ faster than in the infinite-size limit, 
before for $y=t/s\gg 1$ it crosses over to a plateau (full line).
Its height $C_{\infty}^{(2)}$ should scale with $s$ and with $N$. 

The corresponding scaling laws are found 
by repeating the same steps in the calculation of the correlator which led above to (\ref{gl:3.2}).  
Now, we use instead the finite-size scaling function $\mathscr{F}^{(2,N)}$, together with the scaling relation (\ref{gl:3.6}). 
For $t=ys\gg s>0$, the two-time auto-correlator can be written in the form \cite{Henk25e}
\BEA
C\left(ys,s;\vec{0};\frac{1}{N}\right) &=& y^{-\delta} \int_{\mathbb{R}^{2d}} \!\D\vec{U}\D\lap{\vec{U}}\: C_0\bigl(2\lap{\vec{U}} s^{1/2}\bigr)\:
e^{-{\cal M}\bigl(\vec{U}^2+\lap{\vec{U}}^2\bigr)} \mathscr{F}^{(2,N)}\left( \vec{U},\lap{\vec{U}},\frac{\bigl( ys\bigr)^{1/2}}{N}\right)
\nonumber \\
&\simeq& y^{-\delta} \mathscr{F}_C\left( \frac{N}{\bigl( ys\bigr)^{1/2}}\right)
\EEA
where the finite-size scaling form in the second line holds in the scaling limit (\ref{gl:skal}). 
We recover the known result $\delta=\lambda/2$. 

The limit behaviour illustrated in figure~\ref{fig3} fixes the finite-size scaling behaviour of $\mathscr{F}_C$, or equivalently the 
dependence on the third scaling variable of the scaling function $\mathscr{F}^{(2,N)}$. 
Clearly, for $N\gg t^{1/2}=\bigl( ys\bigr)^{1/2}$, the system will behave as being spatially infinite. 
In that limit, $\mathscr{F}_C$ should become a constant and
the scaling function $\mathscr{F}^{(2,N)}\bigl(.,.,\mathfrak{u}\bigr)$ is expected to become independent of $\mathfrak{u}$. 
On the other hand, for finite systems one expects $N\lesssim t^{1/2}$ such that the $y$-independent plateau is reached.
This implies $\mathscr{F}_C(\mathfrak{u})\sim \mathfrak{u}^{-\lambda}$ or equivalently 
$\mathscr{F}^{(2,N)}\bigl(.,.,\mathfrak{u}\bigr)\sim \mathfrak{u}^{\lambda}$. 
Summarising, the plateau height $C_{\infty}^{(2)}=\lim_{s\to\infty} C\bigl(ys,s;\vec{0};\frac{1}{N}\bigr)$ should scale as
\BEQ
C_{\infty}^{(2)} \sim \left( \frac{t}{s}\right)^{-\lambda/2} \left( \frac{t^{1/2}}{N}\right)^{\lambda} \sim N^{-\lambda} s^{\lambda/2}
\EEQ
and in particular, we should have the finite-size scaling behaviour 
\BEQ \label{gl:3.15}
C_{\infty}^{(2)} \sim \left\{
\begin{array}{ll} N^{-\lambda}  & \mbox{\rm if $s$ is kept fixed} \\
                  s^{\lambda/2} & \mbox{\rm if $N$ is kept fixed}
\end{array} \right.
\EEQ
which reproduce \cite{Henk25c} for the special case of quenches to $0<T<T_c$ and for $\mathpzc{z}=2$.

Available tests of this in specific models have been discussed in detail in \cite{Henk25c}. 
There are no known well-studied finite-size effects in the single-time correlator.

{\bf 5.} The global two-time correlator for $t>s$ is obtained by integrating the two-time correlator $C(t,s;\vec{r})$ with respect to $\vec{r}$.
Combining (\ref{corrFT},\ref{eq:2.8},\ref{reponseHE}) leads to \cite{Henk25e} 
\BEA
\lefteqn{ \wht{C}(t,s;\vec{0}) = \int_{\mathbb{R}^d} \!\D\vec{r}\:  C(t,s;\vec{r}) } \label{gl:3.16}
\\
&\simeq&  s^{d/2} \left( \frac{t}{s}\right)^{\Theta} \underbrace{~\int_{\mathbb{R}^{3d}} \!\D\vec{u}\D\vec{U}\D\lap{\vec{U}}\:
C_0\bigl(2\lap{\vec{U}} s^{1/2}\bigr)\: e^{-\frac{\cal M}{2} \vec{u}^2 -{\cal M}\bigl(\vec{U}^2+\lap{U}^2\bigr)}\,
\mathscr{F}^{(2)}\left(\vec{U}+\vec{u}y^{-1/2},\lap{\vec{U}}\right)~}_{=~\mbox{\rm\fns cste.}} ~~~~~
\nonumber \EEA
where in the last line we let $t\gg s$, used as before that $\delta=\xi=-\wit{\xi}$ 
and also the scaling relation (\ref{gl:3.6}) about $\zeta_p$.
As several times before, we also assume that the last integral in (\ref{gl:3.15}) converges to a finite non-zero constant in the $s\gg 1$ limit. 
In particular, the global correlator (\ref{gl:3.16})
with the initial state scales as $\wht{C}(t,0)\sim t^{\Theta}$. Herein, the {\em slip exponent}
\BEQ \label{gl:3.17}
\Theta = \demi \bigl( d - \lambda \bigr)
\EEQ
is given by the extension to $0<T<T_c$ of the Janssen-Schaub-Schmittmann ({\sc jss}) 
critical-point scaling relation \cite{Jans89}, for $\mathpzc{z}=2$, as expected. 
Certainly, the values of $\Theta,\lambda,\mathpzc{z}$ are in general different for $0<T<T_c$ and $T=T_c$.

For quenches onto the critical point $T=T_c$, the original {\sc jss}-relation 
has been the conceptual basis of a whole field of studies on non-equilibrium critical
dynamics, called `short-time dynamics', since it is not necessary to carry out simulation to extremely long times, 
see \cite{Alba11,Zhen98} for classical reviews.
Eq.~(\ref{gl:3.17}) could serve the same purpose in phase-ordering kinetics after a quench into $T<T_c$. 
An example is \cite{Moue25}. 

{\bf 6.} For equal times $t=s$, we might use the combination of (\ref{corrFT},\ref{eq:2.9},\ref{reponseHE}) and find for the squared magnetisation
\BEA
\lefteqn{\hspace{-0.4cm}\left\langle m^2(s)\right\rangle = \wht{C}(s,s;\vec{0}) =\int_{\mathbb{R}^d} e^{-\frac{\cal M}{4}\frac{\vec{r}^2}{s}}
\int_{\mathbb{R}^d} \!\D\lap{\vec{U}}\: C_0\bigl(2\lap{\vec{U}} s^{1/2}\bigr)\: 
e^{-{\cal M}\lap{\vec{U}}^2}\: \mathscr{F}^{(1)}\left( \frac{\vec{r}}{s^{1/2}}\cdot \lap{\vec{U}}\right)}
\nonumber \\
&=& s^{d/2} \underbrace{~\int_{\mathbb{R}^{2d}} \!\D\vec{u}\D\lap{\vec{U}}\: C_0\bigl(2\lap{\vec{U}} s^{1/2}\bigr)\:
\exp\left[-\frac{\cal M}{4} \vec{u}^2 -{\cal M}\lap{\vec{U}}^2\right] \mathscr{F}^{(1)}\left( \vec{u}\cdot\lap{\vec{U}}\right)~}_{=~\mbox{\rm\fns cste.}}
\label{gl:3.18}
\EEA
and with the usual assumption that the last integral converges to a finite, non-zero constant, 
we recover the scaling $\langle m^2(s)\rangle \sim s^{d/2}$ \cite{Jank23}, well-tested in simulations.
Of course, one may obtain this scaling also from (\ref{gl:3.16}) by taking the $t\to s$ limit.

{\bf 7.} To finish, we discuss the finite-size scaling of the global auto-correlator in a fully finite system of linear size $N$.
As above, we expect that the global correlator should converge towards a plateau of height $\wht{C}_{\infty}^{(2)}$
when $\ell(t)\approx N$ but $\ell(s)\ll N$. Generalising (\ref{gl:3.16}) we have, for $t\gg s$ \cite{Henk25e} 
\BEA
\lefteqn{ \wht{C}\left(t,s;\vec{0};\frac{1}{N}\right) = \int_{\mathbb{R}^d} \!\D\vec{r}\: C\left(t,s;\vec{r};\frac{1}{N}\right) }
\nonumber \\
&=& s^{d/2} \left(\frac{t}{s}\right)^{\Theta} \int_{\mathbb{R}^{3d}} \!\D\vec{u}\D\vec{U}\D\lap{\vec{U}}\: C_0\bigl(2\lap{\vec{U}} s^{1/2}\bigr)\:
e^{-\frac{\cal M}{2}\vec{u}^2-{\cal M}\bigl(\vec{U}^2+\lap{U}^2\bigr)}\, 
\mathscr{F}^{(2,N)}\left(\vec{U}+\vec{u}\frac{s}{t},\lap{\vec{U}},\frac{t^{1/2}}{N}\right)
\nonumber \\
&\sim & s^{d/2-\Theta} N^{2\Theta}
\label{gl:3.19}
\EEA
and use of course the scaling relation (\ref{gl:3.17}). The phenomenological discussion of the limits $N\gg t^{1/2}$ and $N\lesssim t^{1/2}$
in the scaling function $\mathscr{F}^{(2,N)}$ and the scaling of the plateau
$\wht{C}_{\infty}^{(2)}$ is as before and leads to the scaling in the last line of (\ref{gl:3.19}). 
The plateau height scales as follows, predicted before for $0<T<T_c$ and $\mathpzc{z}=2$ \cite{Henk25c}
\BEQ \label{gl:3.20}
\wht{C}_{\infty}^{(2)} \sim \left\{
\begin{array}{ll} N^{2\Theta}    & \mbox{\rm if $s$ is kept fixed} \\
                  s^{d/2-\Theta} & \mbox{\rm if $N$ is kept fixed}
\end{array} \right.
\EEQ

\section{Conclusions} \label{sec:con}

The complete known phenomenology of phase-ordering kinetics, after a quench into the phase coexistence region, can be derived from the
covariance of the multi-point response functions under new non-equilibrium representations of the Schr\"odinger Lie algebra \cite{Henk25e}. This reproduces those
properties which are well-established {\it folklore} and also permits to obtain a couple of new scaling laws. We illustrated this here through
a discussion of the two-time and single-time correlation functions. \\

\noindent
{\bf Acknowledgements:} This work was supported by the french ANR-PRME UNIOPEN (ANR-22-CE30-0004-01), 
PHC RILA (51305UC/KP06-Rila/7) and Bulgarian National Science Fund, grant KP-06-N88/3.

\end{document}